\documentstyle[multicol,aps,prl]{revtex}
\begin{document}

\title{Effective Magnetic Hamiltonian and Ginzburg Criterion for Fluids}
\author{Nikolai~V.~Brilliantov$^{1,2}$}
\address{$^{1}$Chemical Physics Theory Group, Department of Chemistry, 
University of Toronto, Toronto, Canada M5S 3H6}
\address{$^{2}$Moscow State University, Physics Department,
Moscow 119899, Russia}
\maketitle

\bigskip
\begin{abstract}

We develop further the approach of Hubbard and Schofield 
(Phys.Lett., {\bf A40} (1972) 245), which maps 
the fluid Hamiltonian onto a magnetic one.  
We show that {\it all} coefficients 
of the resulting effective Landau-Ginzburg-Wilson (LGW) Hamiltonian 
may be expressed in terms 
of the compressibility of a reference fluid containing 
only repulsive interactions, and its density 
derivatives; we calculate the first few coefficients in the case of 
the hard-core reference fluid. From this LGW-Hamiltonian 
we deduce approximate mean-field relations between critical 
parameters and test them on data for  Lennard-Jones,
square-well and hard-core-Yukawa fluids. We 
estimate the Ginzburg criterion for these fluids.

\end{abstract}
\begin{multicols}{2}
 The  modern theory of critical phenomena based on the 
Renormalization Group (RG) technique 
has demonstrated an impressive success in a variety of fields
\cite{goldenfeldma}. However,  most of the studies using 
the RG-approach have been  addressed to 
the criticality of the Ising-like systems. 
Some effective computational schemes based on the RG-technique 
have been developed for fluids \cite{whiteparola}, but they are not 
as convenient for the general analysis of criticality as those based on 
the so-called Ginzburg-Landau-Wilson (LGW) Hamiltonian. 
Effective magnetic Hamiltonians for fluids have been derived 
in a variety of ways. In ref.\cite{sak} the
fluid Hamiltonian was reduced to the magnetic one 
by means of coarse--graining. 
In \cite{evans} the coefficients in an effective
LGW-Hamiltonian were obtained by comparing the  critical amplitudes for 
the order parameter, compressibility, correlation length, etc., calculated 
within generalized
mean-spherical approximation, with those derived from the LGW-Hamiltonian. 
In ~\cite{fishlee} the functional generalization of the Mayer expansion 
for the single--component fluid was used; the attractive interactions were  
treated on the second-virial level, and a few different approximation were 
adopted for the repulsive (hard--core) contribution to the free energy. 
Within these approximations Fisher and Lee evaluated the coefficients for the 
effective LGW-Hamiltonian for a single-component fluid \cite{fishlee};
somewhat different approximations were employed to derive the 
LGW-coefficients for the  restricred primitive model of 
electrolytes \cite{fishlee}.

Hubbard and Schofield  \cite{hubbard} derived the  
effective LGW-Hamiltonians for fluids by an {\it exact} mapping, 
based on the transformation of variables. Although they did not
compute the coefficients of the effective Hamiltonian, they
argued that fluids belong to the Ising universality class. 
In the present study we develop this approach further, 
showing in particular that {\it all} the coefficients of the effective 
LGW-Hamiltonian may be expressed in terms of  the known properties 
of the reference (hard-core) system: the compressibility  and its density 
derivatives. We find explicit  expressions for the first few 
coefficients. Applying the mean-field conditions for the critical point
of the effective magnetic Hamiltonian we formulate simple relations 
between some critical parameters and check them for some fluids; 
for these fluids 
we also estimate the parameter $\tau_G$ of the Ginzburg criterion. 

We start from the fluid Hamiltonian, 

\begin{equation}
H=\sum_{i<j} \phi (r_{ij}) - \sum_{i<j} v(r_{ij}) = H_R +H_{A}
\label{Ham}
\end{equation}
here $\phi(r)$ denotes the  repulsive part of the interparticle 
interaction potential, while $-v(r)$ denotes the attractive part;
$\{ \vec{r}_j \}$ are coordinates of the particles and 
$\vec{r}_{ij}=\vec{r}_i-\vec{r}_j$. 
The attractive part of the Hamiltonian, $H_{A}$,  may be written in terms of 
the Fourier components of the density fluctuations, 
$n_{\vec{k}}=\frac{1}{\sqrt{\Omega}} \sum_{j=1}^{N} e^{-i\vec{k}\vec{r}_j}$, 
and the Fourier transform of the attractive potential, 
$v_k=\int v(r)e^{-i\vec{k}\vec{r}}d\vec{r} $,  as \cite{hubbard,remark119}
\begin{equation}
H_{A}=-\frac{1}{2} \sum_{\vec{k}} v_k n_{\vec{k}} n_{-\vec{k}}+\frac{1}{2}v(0)N ,
\end{equation}  
where $N$ is the number of particles, $\Omega$ is the volume   of the system, 
and summation over the following set of $\vec{k}$ is implied: 
$k_l=\frac{2\pi}{L}n_l$ with $L=\Omega^{1/3}$, 
$l=x,y,z$, and $n_l=0,\pm 1,\pm 2, \ldots$; the thermodynamic limit 
$L \to \infty$ is assumed. Let $\mu$ be the chemical potential of the 
system with the total  Hamiltonian (\ref{Ham}), and $\mu_R$ the 
chemical potential in the reference system having the Hamiltonian 
$H_R$, which includes only repulsive interaction. Then the grand partition 
function, $\Xi$,  may be expressed in terms of that of the 
reference fluid, $\Xi_R$, as \cite{hubbard}

\begin{equation}
\Xi=\Xi_R \left\langle \exp \left\{ \beta \mu' N+\frac{1}{2}\beta 
\sum_{\vec{k}} v_k n_{\vec{k}} n_{-\vec{k}}  \right\} \right\rangle_R 
\label{Xi}
\end{equation}
where 
 $\mu' \equiv \mu-\mu_R+\frac12 v(0)$, and  
$\left\langle \cdots \right\rangle_R$ denotes an  average over 
configurations of the reference system, at the temperature
$T$ with chemical potential $\mu_R$. 
Note that the reference system (with only {\it repulsive} interactions)
does not have a liquid-gas transition; i.e. its grand partition 
function, $\Xi_R$, is regular in the vicinity of the critical point 
of the fluid of interest. 
Following Hubbard and Schofield \cite{hubbard} we use the 
identity 
$\exp(\frac{1}{2} a^2x^2)
=(2 \pi a^2)^{-1/2} \int_{-\infty}^{+\infty} \exp(-\frac{1}{2}y^2/a^2+xy)dy$, 
and after some algebra obtain the ratio $Q=\Xi/\Xi_R$ of the 
grand partition functions \cite{hubbard}:

\begin{eqnarray}
\label{Q2}
&&Q  \propto \int  \prod_{\vec{k}}  d\sigma_{\vec{k}}%
\exp \left\{ \frac{\mu'}{v_0} \Omega^{\frac12}\sigma_0-\frac12 ~\beta^{-1} 
\sum_{\vec{k}} v_k^{-1}\sigma_{\vec{k}}
\sigma_{\vec{-k}} \right\} \nonumber \\
&&\times  \left\langle 
 \exp \left\{ \sum_{\vec{k}}
\sigma_{\vec{k}}n_{-\vec{k}} \right\} \right\rangle _R ;
\end{eqnarray}
here integration is to be performed under the restriction 
$\sigma_{-\vec{k}}=\sigma_{\vec{k}}^*$ 
(the complex conjugate of $\sigma_{\vec{k}}$);  $\sigma_0=\sigma_k$ for $k=0$, 
and we omit a factor regular at the critical point since 
it does not affect the 
subsequent analysis.
Applying the cumulant  theorem ~\cite{cum} to the factor 
$\left\langle \exp \left\{\sum_{\vec{k}} %
\sigma_{\vec{k}}n_{-\vec{k}}\right\}\right\rangle _R$
one obtains ~\cite{hubbard}:

\begin{eqnarray}
\label{Hefgen}
&&Q \propto \int \prod_{\vec{k}}d\sigma_{\vec{k}} 
\exp \left\{-  {\cal H} \right\}, \mbox{~~~with~~}     \\
&& {\cal H}=- h' \sigma_0 \Omega^{\frac12} +     \nonumber \\
&&\sum_{n=2}^{\infty}\frac{1}{\Omega^{\frac{n}{2}-1}}
 \sum_{\vec{k}_1, \ldots, \vec{k}_n} u_{n}'(\vec{k}_1, 
\ldots,\vec{k}_n ) \sigma_{\vec{k}_1} \ldots \sigma_{\vec{k}_n}, \nonumber \\
&&h'=\left( \mu -\mu_R +{v(0)}/{2} \right)v_0^{-1} +\rho  \nonumber \\
&&u_{2}'(\vec{k}_1,\vec{k}_2)=\frac{1}{2!} \delta_{\vec{k}_1+\vec{k}_2,0}%
 \left\{\beta^{-1}v_k^{-1}-
\left\langle n_{\vec{k}_1} n_{-\vec{k}_1} %
\right\rangle _{cR} \right\}~, \nonumber \\
&&u_{n}'(\vec{k}_1, \ldots,\vec{k}_n )=
-\frac{\Omega^{\frac{n}{2}-1}}{n!} \left\langle n_{\vec{k}_1} %
\ldots n_{\vec{k}_n} \right\rangle _{cR},
\mbox{~~~~n $\ge 3$ }\nonumber.
\end{eqnarray}
Here $\rho=N/\Omega$ is the fluid density and 
$\left\langle  \ldots \right\rangle _{cR}$ denotes the {\it cumulant} average
calculated in the reference system. In  Eqs.(\ref{Hefgen}) 
$Q$ has been written in the same way as the partition function for a 
magnetic system having an Ising-like Hamiltonian: 
~$\sigma_{ \vec{k} }$ are the Fourier components of the 
``spin field'', $\sigma( \vec{r} )$, and $h'$ is the ``magnetic field''.

The coefficients of the effective Hamiltonian depend on the correlation 
functions of the reference fluid having only repulsive interactions. 
Using the definitions of the particle correlation 
functions of fluids \cite{gray} 
and definitions of the cumulant averages \cite{cum}, one can {\it directly}
evaluate $\left\langle n_{\vec{k}_1} \ldots 
n_{\vec{k}_n} \right\rangle _{cR}$, and 
thus the coefficients $u_{n}'(\vec{k}_1,\ldots,\vec{k}_n )$. It is 
straightforward to show that $u_{n}'$ may be expressed in terms of the Fourier 
transforms of the correlation functions $h_1$, $h_2 \ldots $, $h_n$ of the 
reference system, defined as 
$h_1(\vec{r}_1)\equiv \delta (\vec{r}_1)$, 
$h_2(\vec{r}_1,\vec{r}_2) \equiv g_2(\vec{r}_1,\vec{r}_2)-1$, 
$h_3(\vec{r}_1,\vec{r}_2,\vec{r}_3) 
\equiv g_3(\vec{r}_1,\vec{r}_2,\vec{r}_3)-g_2(\vec{r}_1,\vec{r}_2)- 
g_2(\vec{r}_1,\vec{r}_3)-g_2(\vec{r}_2,\vec{r}_3)+2$,  etc.
where $g_l(\vec{r}_1, \ldots, \vec{r}_l)$ are $l$-particle correlation 
functions \cite{gray}. In particular, the first few coefficients read:

\begin{eqnarray}
\label{coef1}
&& u_{2}'=  \delta_{1,2} \,\, \rho \left[ \frac{k_BT}{v_0\rho}-
\left( 1+\rho \tilde{h}_2(\vec{k}_1)\right) \right] \\  
&&u_{3}'=-\delta_{1,2,3} \, \, \rho \left\{1+ 
\rho \left[ \tilde{h}_2(\vec{k}_1)+
\tilde{h}_2(\vec{k}_2)+  \tilde{h}_2(\vec{k}_3) \right] 
\right.\nonumber \\
&&\left. +\rho^2 \tilde{h}_3(\vec{k}_1;\vec{k}_2 ) \right\}
\end{eqnarray}
where $\tilde{h}_l$ are the Fourier transforms of $h_l$, and we use the 
shorthand notation: $\delta_{1,2, \ldots, n}=\delta_{\vec{k}_1+
\vec{k}_2 \ldots+\vec{k}_n,0}/n!$.

Now we consider the small-$k$ expansion of the coefficicients. 
First we note that 
the function $\tilde{h}_2(\vec{k})$ may be expressed in terms of the 
Fourier transform of the direct correlation function, 
$\tilde{c}_2(\vec{k})$, as 
$\tilde{h}_2(\vec{k})=\tilde{c}_2(\vec{k})/\left(1-\rho 
\tilde{c}_2(\vec{k}) \right)$, 
and its zero-$k$ value in terms of the isothermal compressibility,
$\chi_{R}=\rho^{-1} \left( \partial \rho / \partial P_{R} \right)_{\beta}$
(where $P_{R}$ is the pressure of the reference fluid) as 
$1+\rho \tilde{h}_2(0)=\rho k_B T \chi_{R} \equiv z_0$.
Using the expansions $v_k=v_0-v_0''k^2+\cdots$ \cite{remark115} and 
$\tilde{c}_2(k)=\tilde{c}_2(0)-\tilde{c}_2(0)''k^2+\cdots$, one obtains 
for $u_2'$, omitting terms of ${\cal O} (k^4)$,

\begin{eqnarray}
\label{u20}
&&u_{2}'= \delta_{1,2} 
\left[a_{2}'+b_{2}' k^2 + \cdots \right]  \nonumber \\
\label{a20}
&&a_{2}'= \left( \beta  v_0 \right)^{-1}-\rho z_0   \\
\label{b2st}
&&b_{2}'=\rho^2 \left[ (z_0^2\tilde{c}_2''(0)+\beta v_0'' 
\left( \rho \beta v_0 \right)^{-2} \right]. 
\end{eqnarray}
The LGW-Hamiltonian does not have terms with powers  
of $k$ higher than $k^2$; moreover, the only term of order 
$k^2$ reads $\propto k^2 \sigma_{\vec{k}}\sigma_{-\vec{k}}$. Thus, 
only zero-order terms should be kept in the expansion  of $u_{n}'$ for 
$n>2$. Hence we may write the contribution of such terms using 
this approximation:

\begin{eqnarray}   
&&u_{3}'=- \delta_{1,2,3}\,\, \rho
\left[1+3\rho \tilde{h}_2(\vec{0})+\rho^2 \tilde{h}_3(\vec{0}) \right] \\
&&u_{4}'=-\delta_{1,2,3,4}\,\, \rho
\left[ 1+7\rho \tilde{h}_2(\vec{0})+6\rho^2 \tilde{h}_3(\vec{0})
+\rho^3 \tilde{h}_4(\vec{0}) \right], \nonumber  
\end{eqnarray}  
etc., where 
$ \tilde{h}_l(\vec{0}) \equiv \tilde{h}_l(0,\ldots,0) $. 
There exists a relation between successive 
correlation functions \cite{gray},

\begin{equation}
\chi \rho^2 
\frac{ \partial}{\partial \rho}  \rho^l g_l = 
\beta \rho^l 
\left[ l\, g_l+\rho \int d \vec{r}_{l+1} \left( g_{l+1}-g_{l} \right) \right],
\end{equation}
from which follows a relation between the functions $\tilde{h}_l (\vec{0})$:

\begin{equation}
\label{relclus}
\chi \rho^2 
\frac{ \partial}{\partial \rho} \rho^l \tilde{h}_l(\vec{0}) = 
\beta \rho^l 
\left[ l\, \tilde{h}_l(\vec{0}) + \rho \tilde{h}_{l+1} (\vec{0}) \right],
\end{equation}
expressing each $\tilde{h}_{l+1} (\vec{0})$ in terms of 
$\tilde{h}_{l}(\vec{0})$ and its density derivative.  
Using Eq.(\ref{relclus}) iteratively one finally finds each 
$\tilde{h}_{l} (\vec{0})$ expressed in terms of the reference system 
compressibility $\chi_R$ and its density derivatives. Explicitly we obtain:

\begin{eqnarray}
\label{u30}
&&u_{3}' \equiv \delta_{1,2,3}\,\,\rho_c u_3= -\delta_{1,2,3} 
\,\, \rho  z_0 \left(z_0 +z_1 \right) \\ 
\label{u40}
&&u_{4}' \equiv \delta_{1,2,3,4}\,\,\rho_c u_4= \nonumber \\
&&=-\delta_{1,2,3,4} \,\, \rho  z_0 \left[z_1^2 + z_0 
\left(z_0+4z_1+z_2 \right)\right],
\end{eqnarray}
defining $u_n$, where $\rho_c$ is the critical density, 
$z_0 \equiv \rho \chi_R/\beta$ as before, 
$z_1=\rho \partial z_0/ \partial \rho$ and 
$z_2=\rho^2 \partial^2 z_0/ \partial \rho^2$. In this way one can evaluate 
{\it all} the coefficients of the effective LGW Hamiltonian of the fluid 
and express them in terms of the compressibility of the reference system 
and its density derivatives. This solves the problem of finding 
the effective LGW Hamiltonian, provided that the compressibility of the 
reference system is sufficiently well known. 

For the reference system with only repulsive interactions one
can often usefully adopt the hard--sphere system with an 
appropriately chosen hard--core diameter \cite{andersen,barker}. 
For soft (not impulsive) repulsive forces a  simple relation ~\cite{barker}
$d=d_{BH}=\int_0^{\sigma} \left[ 1- \exp \left( -\phi(r)/kT \right) \right]$ 
gives the effective diameter $d$ of the hard sphere system
corresponding to a repulsive potential $\phi(r)$ that vanishes at 
$r \ge \sigma$. 
For the hard--sphere system one has the fairly accurate 
Carnahan-Starling (CS) equation of state \cite{barker}, for which 

\begin{equation}
\label{z0}
z_0=\left( 1-\eta \right)^4\left(1+4\eta+4\eta^2-4\eta^3+\eta^4 \right)^{-1},
\end{equation}
where $\eta=\frac{ \pi}{6} d^3 \rho$. The value of $\tilde{c}_2(0)''$ 
may be found from the Wertheim and Thiele solution \cite{remc2} 
for the the direct correlation function, which gives

\begin{equation} 
\label{c2}
\tilde{c}_2(0)''=-\left(\pi  d^5 /120 \right)\,
\left(16-11\eta +4 \eta^2 \right) \left( 1-\eta \right)^{-4}.
\end{equation}
To recast the effective Hamiltonian in the conventional form, we perform 
a transformation from the variables $\sigma_{\vec{k}}$ to 
``field'' variable $\sigma (\vec{r})$. Under this transformation 
integration over the set $\left\{ \sigma_{\vec{k}} \right\}$ becomes
``field'' integration over $\sigma (\vec{r})$, and  
the term $\sim k^2\sigma_{\vec{k}} \sigma_{-\vec{k}}$ 
transforms into $\sim \left( \nabla \sigma (\vec{r}) \right)^2$.
Using $\rho_c^{-1/3}$ as a scaling factor for the length we finally 
arrive at the effective LGW-Hamiltonian:

\begin{equation}
\label{efh2}
{\cal H}=\int d\vec{r}\left[
-h \sigma+ \frac{a_{2}}{2} \sigma^2 
+\frac{u_{3}}{3!}\sigma^3
+\frac{u_{4}}{4!}\sigma^4 \cdots    
+\frac{b_2}{2} \left( \nabla \sigma \right)^2 \right] 
\end{equation}
where 
$h=\left( \mu-\mu_R  +v(0)/2 \right)(v_0\rho_c)^{-1}+(\rho/\rho_c)$, 
$a_2=(k_BT_c/ \rho_c v_0)-z_0 \cdot (\rho/\rho_c)$, and 
$u_{3}$, $u_{4}$ are given by 
Eqs.(\ref{u30}) and (\ref{u40}). The coefficient $b_{2}$ reads:

\begin{equation}
\label{b2}
b_{2}=\frac{1}{80}
\left(\frac{36}{\pi^2 \eta_c}\right)^{\frac13}
\left[ \frac{\lambda_{eff}^2}{\beta \epsilon_{eff}}-B \right]
\end{equation}
where 
$B=4\eta^2 (1-\eta)^4(16-11\eta + 4 \eta^2)/(1+4\eta+4\eta^2-4\eta^3+\eta^4)^2$
and constants $\epsilon_{eff}$ and $\lambda_{eff}$  characterize zero-order, 
$v_0$, and second-order, $v_0''=\frac16 \int r^2 v(r)d\vec{r}$, moments 
of the attractive potential:

\begin{eqnarray}
\label{epsef}
&&\frac{4\pi d^3}{3} \, \epsilon_{eff}
=\int v(r)d\vec{r} = v_0 ~, \\
\label{lamef}
&&\lambda_{eff}^2 d^2=\frac53 \, v_0^{-1} \,\int v(r)r^2 d\vec{r} 
\end{eqnarray} 
This effective LGW-Hamiltonian may be used for analysis 
of the critical behavior, using for example the RG-technique.

Now we perform a mean-field level (MF) analysis based on 
to the effective Hamiltonian. 
To do so we first  remove the cubic term by making the shift, 
$\sigma \to \sigma + \overline{ \sigma}$, with $\overline{ \sigma}$ 
chosen to make the cubic term vanish; this leads to  
new coefficients: $\overline{h}=h+a_2u_3/u_4-u_3^3/3u_4^2$, 
~~$\overline{a}_2=a_2-u_3^2/2u_4$ with $\overline{u}_3=0$,  
$\overline{u}_4=u_4$ and $\overline{b}_2=b_2$.
Then the MF-condition at the critical point, 
$\overline{a}_2=0$ and $\overline{h}=0$, gives the approximate relations

\begin{eqnarray}
\label{a2critcor}
&&\frac{k_BT_c}{\rho_c v_0}= \left[z_0 +\frac{u_3^2}{2u_4} \right]_{c.p.} \\
\label{hcritcor}
&&v_0 \rho_c \left[ 1 + \frac{u_3^3}{6u_4^2}   \right]_{c.p.}
=-\left( \mu-\mu_R  +v(0)/2 \right)_{c.p.}
\end{eqnarray}  
Eq.(\ref{a2critcor}) relates the critical density  
and the critical temperature of the system; 
Eq.(\ref{hcritcor}) relates  the difference of the chemical potentials, 
$\mu-\mu_R $ at the critical point to the critical density. 
We have tested these relations using simulation data for  
Lennard-Jones (LJ) \cite{valLJ}, hard-core-Yukawa (HCY) \cite{lomba}
and square-well (SW) \cite{brilval,vega,analaura} fluids.  
The WCA partition \cite{andersen} of the 
LJ-potential, 
$u_{LJ}(r)=4\epsilon_{LJ} \left[ ( \sigma/r)^{12}-( \sigma/r)^{6} \right]$, 
gives for the attractive part,

\begin{equation}
v(r)=\left\{\begin{tabular}{llll}
$\epsilon_{LJ}$  &~~~~ $r \le 2^{\frac16} \sigma$ \\
$-u_{LJ}(r)$  &~~~~ $ r \ge 2^{\frac16} \sigma $
\end{tabular} \right.
\end{equation}
which is perfectly smooth in the core region. This 
partition gives the best estimates for the thermodynamic functions 
in the WCA--perturbation scheme \cite{andersen}. 
The repulsive part, $\phi(r)=u_{LJ}(r)+v(r)$, is then used to find the 
effective hard-core diameter using the expression quoted above. 
Similarly we use the WCA-partition for 
the SW and HCY fluids. The square-well fluid has the 
interaction potential, $\phi(r)-v(r)=+\infty$ if $r < d$, $-\epsilon$
if $d \le r < \lambda d $ and $0$ if $r \ge \lambda d  $; 
the reference system is the hard sphere system with the diameter $d$. 
We take the attractive part of the potential as  $v(r)=0$ for  
$r \ge \lambda d $, and $v(r)=\epsilon$ 
for $0< r < \lambda d$. 
For the HCY potential, $\phi(r)-v(r)=+\infty$ if $r < d$, and 
$-\epsilon_Y \exp \left[ -\kappa (r-d) \right]/r$ for $r \ge d$;
 the WCA-partition 
gives $v(r)=\epsilon_Y/d$ for $r < d$ and 
$v(r)=\epsilon_Y \exp \left[ -\kappa (r-d) \right]/r$ for $r \ge d$. 

Table I gives the ratio of the right over the left-hand  side of the MF 
Eq.(\ref{a2critcor}) as  $W_c$, and that of Eq.(\ref{hcritcor}) as $Y_c$. 
As one can see from the Table, these MF-relations hold rather 
satisfactorily except for HCY-fluids with short-ranged attractive potentials, 
where the MF-description and MC results apper to differ. 
In Table I we also give some coefficients of the LGW-Hamiltonian at the 
critical point. 

Using coefficients of the effective LGW-Hamiltonian obtained, 
one can also estimate the Ginzburg parameter $\tau_G$ \cite{goldenfeldma}. 
This defines the domain of validity of the classical critical behavior:
the  classical description fails for $ |\tau| \equiv |T/T_c-1|  \ll \tau_G$.
Following  \cite{fishlee}, we write for this parameter:

\begin{equation}
\label{Gi}
\tau_G =\frac{1}{32\, \pi^2} \frac{u_4^2}{ \alpha_2 b_2^3},
\end{equation} 
where $\overline{a}_2=\alpha_2 \tau$. 
From Eqs.(\ref{b2}) and (\ref{Gi}) it follows that for the 
infinitely--ranged Kac--Baker potencial, $\lambda_{eff} \to \infty$, 
(with $\epsilon_{eff} \propto  v_0 $ finite, see Eqs.(\ref{epsef}) and 
(\ref{lamef}) ), 
$b_2 \to \infty$, and thus $\tau_G \to 0$ as expected 
(cf. to \cite{fishlee}).  

To estimate $\alpha_2$ we use the MF-condition (\ref{a2critcor}) 
and approximate  $z_0 \cdot (\rho/\rho_c) +\frac{u_3^2}{2u_4}$ by its value 
at the critical 
point, $\frac{k_BT_c}{\rho_c v_0}$. This yields:

\begin{equation}
\overline{a}_2 
\approx \frac{k_BT}{\rho_c v_0}-\frac{k_BT_c}{\rho_c v_0}=\alpha_2 \tau
\end{equation}
Thus, $\alpha_2=\frac{k_BT_c}{\rho_c v_0}=k_BT_c/8 \eta_c \epsilon_{eff}$. 
Using this value of $\alpha_2$, and $u_4$,  $b_2$  given by 
Eqs.(\ref{u40}), and (\ref{b2})
(the coefficients in this relation computed at the critical point), 
we calculate an approximate Ginzburg parameter $\tau_G$ for some of the SW, 
HCY and LJ-fluids. 
The results are given in Table I \cite{remrig}. 

As we see in Table I, the derived value of $\tau_G$ is of the order 
of $\approx 10^{-1}$ for the most of fluids studied in computer 
simulations and lies within the range of $\tau_G$ values 
predicted in Ref.\cite{fishlee}. 
On the other hand, 
for the SW-fluid with the most long-ranged attraction 
($\lambda_{eff}=3$), $\tau_G \approx 10^{-2}$, is much smaller, 
and may explain the MF-like behavior 
observed  \cite{analaura}. For the LJ-fluid, with an 
attractive potential, $\propto r^{-6}$, similar to that of real  fluids,
$\tau_G$ is of the order of $10^{-2}$; 
real simple fluids are supposed to have similar 
values of $\tau_G$  \cite{anis}. 
However, it may be noted that in 
MC simulations of the LJ-fluid \cite{valLJ} (where a  
cut-off of the LJ-potential and tail corrections were used)  
Ising-like behavior was observed to values of $ |\tau|$, at least few times
larger than the $\tau_G$ given in Table I. 

\bigskip
Valuable discussions with John Valleau, his help and suggestions
are highly appreciated. Financial support of NSERC of 
Canada is acknowledged.


{\bf Table I} Gives the Ginzgurg criterion, $\tau_G$,  
the ratio of the right over the left-hand  side of 
Eq.(\ref{a2critcor}) as  $W_c$, and that of Eq.(\ref{hcritcor}) as $Y_c$, 
and  coefficients of the effective LGW-Hamiltonian, $b_{2,c}$, 
$u_{4,c}$, $\alpha_2$. Here 
$\lambda_{eff}$ is the effective range of the attractive potential 
($\lambda_{eff}=\lambda$ for the SW-fluid), $T^*=k_BT/\epsilon_{eff}$ 
and $\rho_c^*=\rho_c d^3$ are respectively, the reduced critical temperature 
and density. 
The critical data for SW-fluid is from \cite{vega}, from 
\cite{brilval} (for $\lambda_{eff}=1.5$), from 
\cite{analaura} (for $\lambda_{eff}=3.0$); for the HCY-fluid from 
\cite{lomba}; for the LJ-fluid from \cite{valLJ}.

\begin{tabular}{||c|c|c|c|c|c|c|c|c||}\hline \hline
$\lambda_{eff}$ & {$T_c^*$}  & $\rho_c^*$ & $ \tau_{G}$  &
$W_c $ & $Y_c $ & $b_{2,c}$ & $u_{4,c}$ & $\alpha_2$ \\ 
{}&{}&{}&{}&{}&{}& $\times 10^2 $ & $\times 10^2$  & {$\times 10^2$ }\\ \hline
    	{}     &      {}    &     {}     &  {}     & {SW}  &  {}     &  {}      &   {}    & {}    \\  
   1.25        &     0.391  &    0.370   & 0.272   & 1.07  &  {}     &  1.21    & 0.62    & 25.2 \\  
   1.375       &     0.375  &    0.355   & 0.259   & 1.06  &  {}     &  1.55    & 0.88    & 25.2 \\  
   1.500       &     0.370  &    0.329   & 0.301   & 1.06  &  1.01   &  2.00    & 1.46    & 26.9 \\  
   1.75        &     0.338  &    0.284   & 0.407   & 1.12  &  {}     &  2.89    & 2.96    & 28.4 \\  
   2.00        &     0.346  &    0.225   & 0.334   & 1.09  &  {}     &  4.58    & 6.10    & 36.7 \\  
   3.00        &     0.433  &    0.181   & 0.012   & 0.87  &  {}     & 15.7     & 9.02    & 57.0 \\  
        {}     &     {}     &            & {}      & {HCY} &  {}     &   {}     & {}      &  {}   \\  
   1.396       &     0.298  &    0.377   & 0.336   & 1.45  &  0.56   &  1.91    & 0.52    & 18.8 \\  
   1.579       &     0.307  &    0.375   & 0.096   & 1.39  &  0.62   &  1.70    & 0.55    & 19.5 \\  
   2.137       &     0.328  &    0.313   & 0.054   & 1.16  &  0.95   &  4.40    & 1.91    & 25.0 \\  
     {}        &     {}     &            & {}      & {LJ}  &  {}     &    {}    &   {}    & {}    \\  
   2.215       &     0.347  &    0.320   & 0.027   & 1.06  &  0.94   &  5.09    & 1.70    & 25.9 \\  
\hline
\hline
\end{tabular} 

\end{multicols}


\begin{references}
\bibitem{remark119}One can always choose $v(r)$ with $v(0)$ being finite.
\bibitem{remark115}
	Here we restrict ourselves to potentials decaying 
	at $r \to \infty$ faster than $r^{-5}$ (in three dimensions).
	For such potentials the expansion, 
	$v_k=v_0-v_0''k^2+\cdots$, is valid 
	and one can construct the effective LGW-Hamiltonian.
\end{references}
\end{document}